Action at a distance in Quantum Theory
Jerome Blackman

This paper started as an attempt to create a consistent model for the puzzling "action at a distance" problem in quantum theory that was raised as a consequence of the famous paper by Einstein, Podolsky and Rosen [1]. Actually, in that paper the authors were trying to show that the eigenvalues of two non-commuting observables could be simultaneously measured but the argument depended on action a distance being impossible, but at present the EPR problem is generally taken as referring to how two systems, ostensibly separated by some distance, can interact in opposition to the limits imposed by special relativity. This difficulty is usually glossed over by the not very satisfactory observation that this does not allow information to be transmitted faster than light.

Suppose we start with the 2 electron problem, i.e. the case where one electron has spin "up" and the other spin "down" They are brought together so that their spins interact and then separated to some arbitrary distance. It is then known, both experimentally and from the conservation of spin, that if one is measured with spin up the other will always be measured with spin down. The first problem that arises in trying to analyze this problem is to find the proper Hilbert space representation of two electrons interacting and, since special relativity may be involved, it should appear in an appropriately relativistic setting. The usual way to do this in the non relativistic case is to take the Hilbert spaces of the two electrons $\mathcal{H}_i$ for i=1,2 and form their tensor product $\mathcal{H}_1 \otimes \mathcal{H}_2$. In the non-relativistic case there is no problem but if one tries to generalize to the relativistic case with each electron having its own time there seems no way to go further [2,Chapter ], We will start with the non relativistic view and we will assume that the time development of the state vector in $\mathcal{H}_1 \otimes \mathcal{H}_2$ is governed by a unitary group $U(\tau)$. The point of view is that the basic structure should be based on the quantum mechanics paradigm and that the relativistic structure should be derived from that. In particular we want to show how the three or four dimensional metric or pseudo metric structures of Newton or Einstein are related to that of Hilbert space. We start by looking at the problem from the view of a single observer and will later show how other observers can be introduced in a way appropriate for special relativity . It would be nice to have a completely rigorous presentation of the theory but there are gaps that are bridged as best we can. The theory does however shed light on the EPR problem and perhaps a somewhat different view of the relation between quantum theory and relativity.

For a discussion of this problem it as appropriate to consider the more general problem of a large number of particles so we will look at

$$\mathcal{H} = \mathcal{H}_1 \otimes \mathcal{H}_2 \otimes ... \mathcal{H}_n .$$  (1)

We take a representation of $\mathcal{H}$ which is a slight generalization of that implied by the Schrödinger equation. Let $S$ be a measure space with measure $\mu$. Although this may appear to be an unnecessary generalization $S$, with additional imposed conditions, will become either the three dimensional space of Newtonian physics or, with time added, the pseudo Euclidean space of Einstein.

On $S$ we can construct the Hilbert space of all square integrable complex functions with inner product $\langle f|g\rangle = \int_S \overline{f}g d\mu$. We need $n$ copies of this space Call these Hilbert spaces $L_i^2$. This type of Hilbert space needs an added term to account for possible internal degrees of freedom such as charge or spin. Call it $\mathcal{G}$. Then define the representation of $\mathcal{H}_i$ to be $L_i^2 \otimes \mathcal{G}_i$. Then a point in the representation of $\mathcal{H}$ is a direct sum of terms of the form

$$f_1 \otimes g_1 \otimes f_2 \otimes g_2 \otimes ... \otimes f_n \otimes g_n \tag{2}$$

where $f_i \in L_i^2$ and $g_i \in \mathcal{G}_i$. If the particles are non interacting the general state would be a direct sum

$$\oplus_i f_i \otimes g_i \tag{3}$$

In order to motivate the next step let us jump ahead and assume that $S$ is a three dimensional Euclidean space $E$ and let's consider a case where $\mathcal{H}$ is of the form $\mathcal{H} = \mathcal{H}_1 \otimes \mathcal{H}_2 \otimes \mathcal{H}_m$ where $\mathcal{H}_1$ and $\mathcal{H}_2$ are the state spaces for two independent particles and $\mathcal{H}_m$ is the state space of a measuring device which we will use to locate the positions in the Euclidean space of the two particles.

For $\mathcal{H}_i$ with $i = 1, 2$ we take the representation $L_i^2$ dropping the $\mathcal{G}_i$ term since we are only interested here in a position measurement. The general theory of measuring an observable with a finite spectrum is presented in [3] along with an outline of the extension to infinite spectrum cases. The infinite spectrum case for the classical wave particle duality problem is treated in [4] but for our purposes here we will only state the result we need. The discussion in [4] covers the position measurement of the particle as described in the two slit experiment so it only measures two of the coordinates rather than three but the mathematics of a generalization to the third would be clear even if the actual design of the experiment might be difficult. If we try to measure the position of the first particle it can be accomplished by a unitary transformation in the space $L_1^2 \otimes \mathcal{H}_m$ and will result in state vector in $L_1^2$. If the position operator had a discrete spectrum the function would be an eigenfunction but as shown in [4] since the spectrum is continuous it will be a function whose support lies in the space occupied by the individual detectors of the measuring device. All we need here is the fact that it is a vector in $L_1^2$. Similarly if we were to measure the position of the second particle we would find that it is vector in $L_2^2$. On the other hand, to the physicist doing the measuring both particles are in the same space namely the ordinary three dimensional space we live in. Now a point in $L_i^2$ is a function with very localized support in $E$ giving the position in our ordinary space so it is reasonable to hypothesize that $E$ is what we call our space (at least for the single observer we have). Indeed, we can go further to consider the case where each of the $L_i^2$ is the space of a macroscopic object, say a stone, and the stones are not interacting. Then as is well known the Schrödinger equation will predict that the stones will behave dynamically the same way as predicted by Newtonian mechanics. Of course determination of other properties of the stone would involve interaction with its state vector in its Hilbert space.

The epistemological view embraced by most (but not all) scientists is that there is a 'real' world existing independently of our consciousness and that using the chemical and electric signals we receive in our brains through our senses we construct models that we use to survive and do physics. We can then conjecture that sometime early in the development of central nervous systems sentient creatures developed the three dimensional model of $\mathcal{H}$. In modern terminology it was perhaps the first 'effective' theory in physics.

At this point it is necessary to explain where the metric of $S$ comes from and show how special relativity enters the picture. It is also the point where the incompleteness of the theory becomes apparent, leaving a gap that we hope can be filled in the future. What we need is the existence of the electromagnetic field in $S$ along with Maxwell's equations so we would like to be able to show how the existence of photons in $\mathcal{H}$ can be "projected" onto $S$ along with a unitary group action leading to these equations. This is not far from the usual interpretation of quantum theory so we will assume it can be done.

Now consider our observer who is basically described as a state vector in $\mathcal{H}$, but who considers himself as a projection in $S$. Due to the electromagnetic field which governs the binding of chemicals he now has access to measuring sticks from which he can deduce a metric. Based on the properties revealed by these sticks Euclid created his geometry. "Errors" were made of course. Since the color of light was detected though the eyes and radiant heat through the skin these were not recognized as the same phenomenon until the nineteenth century. Also since the time development of events in $\mathcal{H}$ is governed by the unitary group $U(\tau)$ it is natural for him to take his time $t$ as being the same as $\tau$. But if the metric adopted by our first observer is based on the properties of light which has the same speed to all other observers moving at a constant velocity with respect to the first it becomes clear that the two observers have different metrics in their version of $S$ and therefore in their construction of $E$. This was all worked out by Einstein more than a hundred years ago. In particular the time $t'$ of the second observer is related to the time $t$ of the first observer by the equation

$$dt' = dt\sqrt{1-v^2} = \tilde{v}dt \text{ where } \tilde{v} = \sqrt{1-v^2}. \qquad (4)$$

Since $v$ is constant the two times are related by

$$t' = \tilde{v}t + k \text{ where } k \text{ is a constant.} \qquad (5)$$

Therefore the unitary group in terms of the second observer's time is given by

$$U(t') = U(\tilde{v}t + k) = U(\tilde{v}t)U(k) \qquad (6)$$

It is easy to see that if $U(t)$ is unitary so is $U(t')$. $U(k)$ acts as a translation in time or a resetting of the clock and can be ignored. It follows from this construction that both observers will see the action of $U$ the same way but on a different time scale and that there will be no preferred inertial observer. It is easy to show that the contraction in length in the direction of motion of one observer relative to the other follows from the contraction in time given by (5) so this picture is consistent with special relativity and says that each inertial observer has the same laws of quantum theory as any other.

Observe also that if a problem can be resolved completely within the large Hilbert space $\mathcal{H}$ then it will be resolved for every inertial observer automatically. The theory of measurement presented in [3] and [4] is handled that way. Unfortunately the tools for calculations in $\mathcal{H}$ are scarce as we shall see below.

If we think of a light cone diagram with the observer suddenly changing velocity at a certain time then his clock will measure time in correspondence with the new velocity and since any reasonable curve in the light cone can be approximated by a sequence of straight lines it follows that the laws of quantum theory for such an observer is measured by his proper time.

We can now return to the two electron problem. Initially the electrons are not interacting so each lies in its own Hilbert space $\mathcal{H}_i$ which we take as $L_i^2 \otimes \mathcal{G}_i$ for $i = 1, 2$. Here $\mathcal{G}_i$ is a two dimensional sin space with basis $|\uparrow\rangle$ and $|\downarrow\rangle$. The two state vectors are then of the form $f_i(\vec{x}) \otimes s_i$ where $s_i \in \mathcal{G}_i$. While the complete formalism for describing the action of bringing the two electrons together to intertwine their spin and then separating them is unknown we can proceed with the assumption that they end up separated in space but with spin intertwined. Steven Weinberg [5] calls this a one particle state (but not an elementary particle). It lies in $L_1 \otimes L_2 \otimes \mathcal{G}_1 \otimes \mathcal{G}_2$. We are assuming that the electrons (aside from their spin) are not intertwined so they can be again be indicated by the subscripts 1 and 2 although perhaps not by the same functions. Because electrons are fermions the effective spin space is two dimensional with basis $|\uparrow, \downarrow\rangle$ and $|\downarrow, \uparrow\rangle$. The representation of the particle would then be of the form

$$f'_1 \otimes f'_2 \otimes (a|\uparrow, \downarrow\rangle + b|\downarrow, \uparrow\rangle). \tag{7}$$

where $|a|^2 + |b|^2 = 1$. According to the classical Bohr theory of measurement or the one appearing in [3] a measurement of spin will either yield $|\uparrow, \downarrow\rangle$ with probability $|a|^2$ or $|\downarrow, \uparrow\rangle$ with probability $|b|^2$. In any case it would be sufficient to measure only the spin of one of the electrons.

This example raises another question which we will return to below but now let's turn to another action a distance problem, namely the splitting of the wave function of a particle to large distances and then the apparent fact that the detection of the particle at one location seems to make the other part of the wave function disappear "instantly". This is a particularly simple problem in the framework we are using since only one particle is involved. Since we are not dealing with spin we can take the representation of the Hilbert space as just $L^2$. Then the state vector will be of the form $f_1(\vec{x}) + f_2(\vec{x})$ where the support of the $f_i$ have an empty intersection and the sum of their squares is 1. The general theory as described in [3] describes the measurement as being accomplished by a unitary group of transformations in the space $L^2 \otimes \mathcal{H}_m$ where $\mathcal{H}_m$ is the Hilbert space of the measuring device. The experimenter may think he is discovering whether the electron is located in the support set of $f_1(\vec{x})$ because he is thinking in the Euclidean space but in the Hilbert space the operation is on the whole vector $f_1(\vec{x}) + f_2(\vec{x})$. Since the measurement is carried out by a unitary transformation if the transformation takes $\vec{f_1}(x)$ into 0 it will take $\vec{f_2}(x)$ into a vector with norm 1. The procedure is explained in more detail in [4].

The whole action at a distance problem seems to stem from confounding two different structures. One is the Hilbert space structure of quantum theory and the other is the metric space that we are used to. Hilbert space has an inner product but not a distance. The metric structure comes from the electromagnetic field and it is different for observers in different inertial frames. Since the electromagnetic field is now unified with the weak and strong force it is reasonable to believe that these fields can also use a metric space as background although it is necessary to add dimensions to accommodate various charges. One of the main causes of the confusion it seems is that for many problems the $L^2$ representation is too close to the metric space idea. The representation of the Hilbert space consists of functions and their vector space properties. The space on which the functions are defined is something quite different and the metric on that space does not carry over to the Hilbert space. Early in the development of quantum theory there were attempts to find direct evidence of the wave function in space based on that confusion.

Furthermore while we have shown how the ordinary space for a collection of independent particles or objects can be realized as a simplification of a particular representation of the Hilbert space that does not carry over in any obvious way to even two particles if they are interacting.

Returning to the two electron problem above it is known that in spite of the "instantaneous" transmission of spin states the phenomenon can not be used to transmit information at speeds greater than the speed of light. That fact rests on Bohr's statistics for the result of measuring spin which is stated above. However since in [3] the reason for Bohr statistics is given the question can be reopened. It turns out that if the device measuring spin could be modified in a certain way and if the theory is correct then the linking of the spin might make it possible to transmit a signal faster than the sped of light.

The measurement theory of [3] considers a measurement of an observable when the observable has a finite number of eigenvalues $|p_i\rangle$ to be a unitary group in the space $\mathcal{H}_m \otimes \mathcal{H}_p$ where $\mathcal{H}_m$ is the Hilbert space of the measuring device and $\mathcal{H}_p$ is the space of the particle. There are two conditions required for the unitary group to be a measurement. The first is just the requirement that the final state of the measuring device be different for each eigenvector and the second that

$$H(\mathcal{H}_m \otimes \mathcal{H}_i) \subseteq \mathcal{H}_m \otimes \mathcal{H}_i \text{ for each i.} \qquad (8)$$

In this equation $H$ is the Hamiltonian of the unitary group and $\mathcal{H}_i$ is the one dimensional Hilbert space of $|p_i\rangle$. This is just the statement that if the particle is in the $i^{th}$ state the measurement doesn't introduce any other components. In the simple case of the two electron the action of the unitary group has the effect of equation [8] having $a$ and $b$ being tensors in $\mathcal{H}_m$ which are also functions of the group parameter $\tau$. The equation $|a|^2 + |b|^2 = 1$ still holds now referring to Hilbert space norm rather than absolute value. Since $|a|$ and $|b|$ are functions of $\tau$ the number $|a|$ can be considered to describe a random walk on the unit interval because of the action of the measuring device which is assumed to consist of many particles with indeterminate motion. In [3] the continuous process is approximated by a discrete process which transforms the problem into a random walk. The probability of the motion is assumed to be balanced i.e. a step to the right of to the left is assumed to be equal to $1/2$ which yields the Bohr statistic. If one could arrange the measuring device so that it could change the probability at will so that the probability of going to the left or tight was greater than $1/2$ a signal could be sent faster than the speed of light. Equation (8) in this simple case merely says that the points 0 and 1 are absorbing barriers so that if the point reaches either end of the interval in the process of the measurement it stays there. It isn't clear that this is possible but experimental physicists are a very ingenious lot.

There are several interesting problems that this paper raises:

1. What is its relation to QFT?

2. Equation (4) is really an equation in a tangent space. It yields the familiar Minkowski space but if one wanted to consolidate the individual spaces of the different observers taking the necessary second order corrections to (4) into account the results might be interesting.

3. If the origin of the metric structure of ordinary space given above is taken seriously then it raises the question of whether even the metric of general relativity in the first seconds after the big bang can be trusted.

The above theory may or may not be useful but the existence of action at a distance seems to be a signal that nature is trying to tell us something.

References

[1] Einstein, A., Podolsky, P., and Rosen, N. Can quantum-mechanical description of physical reality be considered complete? Phys. Rev. 47 777-780 (1935)
[2] Penrose, R. The Road to Reality, Alfred A Knopf, New York (2006)
[3] Blackman, J., and Hsiang, W.T. Why Probability Appears in Quantum Theory. Phys. Essays 26,1 34-39 (2013)
[4] Blackman J., On Wave Particle Duality. Phys Essays 26, 3 347-349 (2013)